
\magnification = 1200
\hsize = 6 true in
\vsize = 8.5 true in
\def \n {{\bf n}}
\overfullrule = 0 pt
\font\ninerm = cmr9
\centerline {COSMOLOGICAL ``GROUND STATE'' WAVE FUNCTIONS IN GRAVITY}
\centerline {AND ELECTROMAGNETISM}
\vskip 0.5 true in
\centerline {Michael P. Ryan, Jr.}
\centerline {{\it Instituto de Ciencias Nucleares}}
\centerline {{\it Universidad Nacional
Aut\'onoma de M\'exico}}
\centerline {{\it A. Postal 70-543, M\'exico 04510 D.F.}}
\centerline {{\it MEXICO\/}}
\vskip 1 true cm
\centerline {ABSTRACT}
\vskip 10 pt
{\leftskip 3 pc \rightskip 3 pc \ninerm \baselineskip 10 pt
The coincidence of quantum cosmology solutions generated by solving a
Euclidean version of the Hamilton-Jacobi equation for gravity and by
using the complex canonical transformation of the Ashtekar variables
is discussed.  An examination of similar solutions for the free
electromagnetic field shows that this coincidence is an artifact of
the homogeneity of the cosmological space.
\smallskip}
\vskip 7 pt

\noindent {\bf 1. Introduction}
\vskip 10 pt

Jerzy Pleba\'nski is one of the pioneers in complex methods applied to
general relativity, especially in the field of self-dual solutions.
It is one of the ironies that seem to abound in physics that the germ
of what are now called the Ashtekar variables is contained in a 1977
paper of Jerzy's$^1$.  If the Hamiltonian form of the self dual
action presented there had been constructed, we might today be using
the name ``Pleba\'nski variables'' instead of Ashtekar variables.  In
any case, everyone who works, no matter how briefly or superficially,
in complex relativity will find himself acknowledging Jerzy's work.

It is particularly fitting that those of us who work in relativity in
Mexico acknowledge Jerzy's contribution in reviving the study of
gravitation in this country (One must say ``revive'' because almost
fifty years ago Marcos Moshinsky, Alberto Barajas and Carlos Graef were
making original contributions to the field$^2$.) and creating a climate
in which the study of gravitation was considered an interesting and
valuable endeavor.

The present article considers the application of the Ashtekar
variables$^3$ to finding exact solutions in quantum gravity, especially
minisuperspace or quantum cosmology solutions.  The study of quantum
minisuperspaces has a relatively long history since they were
introduced by Charles Misner in the late sixties, and they
have often been used as models for the quantization of gravity as well
as being interpreted, especially by Stephen Hawking and his
collaborators, as giving information about early epochs of the
Universe.  Here I will focus on quantum cosmologies
as models for the use of Ashtekar variables in quantum gravity.

A number of people have used the Ashtekar variables to find solutions
to quantum cosmologies, including Ashtekar himself, Pullin$^4$,
Obregon$^5$, and myself and Moncrief $^6$.  As will be mentioned below, one
usually splits the cosmological minisuperspaces into a
non-dynamical, time-independent part and a finite or infinite set of
time-dependent variables that define the state space of the
minisuperspace.  The constraints of gravity (ADM or Ashtekar)
can then be expressed in terms of these variables and a quantum
cosmology can be constructed by turning the variables and their
conjugate momenta into operators and applying them to a state
function that is a function of some of them.  In the ADM formalism
the operator corresponding to the Hamiltonian constraint leads to the
Wheeler-DeWitt equation and in the Ashtekar formalism the equivalent
constraints lead to what has been called the Ashtekar-Wheeler-DeWitt
equation$^6$. The Ashtekar-Wheeler-DeWitt equation is closer in
spirit to the Bargmann$^7$ formalism in ordinary quantum mechanics, so
the state function that is obtained as a solution must be interpreted
in a different way from the usual ADM wave function in that a norm
over these functions must be constructed or they must be mapped
(perhaps by means of a kernel such as that constructed by Bargmann
for the harmonic oscillator$^7$) into an ordinary ADM wave function.

The existence of such mappings allows one to find solutions to the
Ashtekar-Wheeler-DeWitt equation and then map them to solutions of
the usual Wheeler-DeWitt equation.  For a simple-minded mapping this
was done in Ref. [6], where a constant solution to the
Ashtekar-Wheeler-DeWitt equation for a diagonal Bianchi Type IX
cosmological model was mapped to a ``ground state'' solution of the
Wheeler-DeWitt equation for a certain factor ordering that had the
form $k\times e^{-S}$, $k =$ const., where $S$ was a solution of the
Einstein-Hamilton-Jacobi equation$^8$.  This technique was extended to
a supersymmetric theory by Obregon, Pullin, and myself~ $^5$.

This procedure might seem to be a ``magic bullet'' that would allow
one to find solutions of the $e^{-S}$ form for many metrics in
quantum gravity, not only the quantum cosmologies.  I plan to show
here that while the solutions to the Wheeler-DeWitt equation obtained
by mapping constant solutions of the Ashtekar-Wheeler-DeWitt equation
always exist, they are not necessarily ``ground state'' solutions
in the sense of Ref. [6].
I will give a few examples and study the Maxwell analogue of this
procedure where the difference between the mapped ``Ashtekar''
solutions and the $e^{-S}$ solution is more obvious.

The plan of the paper is as follows.  In Sec. 2 I will give a sketch
of the Bargmann representation for the harmonic oscillator, along
with a modified
representation that is closer in spirit to the
Ashtekar representation. In  both cases I will give mappings between
the Bargmann or ``Ashtekar'' state and states in the ordinary
coordinate representation.  In Sec. 3 I will give several quantum
cosmology solutions obtained by the mapping procedure, some of which
are more useful than others.  Section 4 is devoted to the
electromagnetic analogue of the procedure and Sec. 5 contains
conclusions.
\vskip 20 pt
\noindent {\bf 2. The Bargmann Representation for the Harmonic
Oscillator}
\vskip 10 pt

In 1961 Bargmann$^7$ studied in great detail a representation
introduced originally by Fock$^9$ in the quantum mechanics of the
harmonic oscillator.  The basic idea is the simple-minded one of
taking the annihilation and creation operators  $\hat a = (1/\sqrt
{2})(q - ip)$, $\hat a^{\dagger} = (1/\sqrt{2})(q + ip)$ for the
harmonic oscillator and achieving the commutation relation $[\hat
a^{\dagger} , \hat a ] = 1$ by realizing $\hat a^{\dagger}$ as $\partial
/\partial a$.  While this concept is seductive, $\hat a$ and $\hat
a^{\dagger}$ are not unitary operators and the space of
functions ${\cal
F}(a)$ over the complex numbers $a = (1/\sqrt{2})(q-ip)$ upon which
the operator $\partial /\partial a$ is supposed to act is not a
Hilbert space.  Bargmann's contribution was to make sense of this
idea by means of a careful analysis of the problem.  An important part
of his
contribution was to construct a norm on the function space
${\cal F}(a)$ and to give a mapping from ${\cal F} (a)$ to the
ordinary Hilbert
space of the harmonic-oscillator coordinate wave
functions.

Bargmann defined the norm on ${\cal F} (a)$ for an $n$-dimensional
phase space.  It is sufficient to show his result for a
two-dimensional phase space $(q, p)$ since the overall structure is
the same.  He defines the inner product of two functions $\psi_1 (a)$
and $\psi_2 (a)$ in ${\cal F} (a)$ by means of a kernel $\rho (q,p)$
as
$$ (\psi_1 ,\psi_2) = \int \psi_1^* \psi_2 \rho (q,p) \, dqdp. \eqno
(2.1)$$
He then determines $\rho$ by demanding that $\partial /\partial a$ be
the adjoint of $a$ with respect to this inner product, which leads
to a set of partial differential equations for $\rho$.  He finds
$\rho = c \exp(-a^* a)$ = $c\exp (-[p^2 + q^2])$.

He next gives the mapping from the Hilbert space of the ordinary
harmonic oscillator wave functions, also in terms of a kernel $A
(a,q)$.  This mapping from a wave function $\psi (q)$ to a wave
function $\Psi (a)$ in ${\cal F} (a)$ has the form
$$\Psi (a) = \int A(a,q) \psi (q) \, dq.  \eqno (2.2)$$
Demanding that $\hat a =(1/\sqrt{2})(q - \partial /\partial q)$ and
$\hat a^{\dagger} = (1/\sqrt{2})(q + \partial/ \partial q)$ map into
$a\Psi (a)$ and $\partial \Psi (a)/\partial a$, respectively, again
implies a set of partial differential equations for $A$.  These
equations have as a solution
$$A(a,q) = {\pi}^{-1/4} \exp [-{{1}\over {2}}(a^2 + q^2) - \sqrt{2}
aq],  \eqno (2.3)$$

If one now writes the time-independent harmonic-oscillator wave
equation in terms of the  operators $a$ and $\partial /\partial a$,
one finds that, for the usual factor ordering,
$$\left (a{{\partial}\over {\partial a}} + {{1}\over {2}} \right )\Psi (a) =
E\Psi (a). \eqno (2.4)$$
Here we see the advantages of ``Bargmannization'', since this
equation is much easier to solve than the usual Schr\"odinger
equation for the harmonic oscillator.  The general solution is $\Psi
(a) = c a^{E-{{1}\over {2}}}$, $c =$ const.  Bargmann gives the
simplest orthonormal (with respect to the inner product [2.1]) set of
eigenstates as $u_n = a^n /\sqrt {n!}$.  These states correspond to
energy eigenstates with $E = n + {{1}\over {2}}$.

Bargmann also shows that the kernel (2.3) maps the eigenstates $u_n$ to
the usual harmonic oscillator energy eigenstates $\phi_n = [2^n n!
\sqrt {\pi}]^{-1/2} H_n (q)
e^{-q^2/2}$.  Of particular interest is the ground state.  The
ground state in the Bargmann representation is $u_0 = 1$.  This
constant wave function maps to $\phi_0 (q) = (\sqrt {\pi})^{-1/2} e^{-q^2/2}$.

The Ashtekar variables are not exact analogues of the Bargmann
variables for the harmonic oscillator.  They are closer in spirit to
the following transformation
$$ Q = q, \qquad P = q + ip. \eqno (2.5)$$
As before, we would like to achieve the commutation relations $[\hat
P, \hat Q] = 1$ by realizing $\hat P$ as $\partial /\partial Q$.
Again we can follow Bargmann's procedure, finding a kernel $\rho (Q)$
that gives an inner product on the function space ${\cal F} (Q)$, and
another kernel $A(q, Q)$ that maps the ordinary harmonic-oscillator
wave functions $\psi (q)$ to the ``Ashtekar-like'' functions $\Psi
(Q)$.  The kernel $\rho$ can be found by requiring that $\hat
P^{\dagger} = \hat q - i\hat p = -\hat P + 2\hat Q$ be the adjoint of
$\partial /\partial Q$, that is, if $(\psi_1, \psi_2) = \int \psi^{*}
(Q) \rho (Q) \psi_2(Q) \, dQ$,
$$ (\psi_1, \partial \psi_2 /\partial Q ) = ([-\hat P + 2Q]\psi_1 ,
\psi_2), \eqno (2.6)$$
or,
$$ \int \rho (Q) \psi_1^{*} {{\partial \psi_2}\over {\partial Q}} \,
dQ = \int \left (-{{\partial \rho}\over {\partial Q}} \psi_1^{*}
\psi_2 - \rho {{\partial \psi_1^{*}}\over {\partial Q}} \psi_2 \right
)\, dQ$$
$$= \int \rho (Q) \left (-{{\partial \psi_1^{*}}\over {\partial
Q}} \psi_2 + 2Q\psi_1^{*} \psi_2 \right ) . \eqno (2.7)$$
These relations imply $-\partial \rho/\partial Q = 2Q\rho$, or $\rho =
e^{-Q^2}$.

For the kernel $A(q,Q)$ use Bargmann's condition that $Q\Psi$ be the
transform of $q\psi$ and that $\partial \Psi/\partial Q$ be the
transform of $(q + i\hat p)\psi$.  These two conditions are
$$ Q\Psi = \int QA(q,Q) \psi(q) \, dq = \int A(q,Q) q\psi (q), \eqno
(2.8)$$
$${{\partial \Psi}\over {\partial Q}} = \int {{\partial A}\over
{\partial Q}} \psi \, dq = \int A\left ({{\partial \psi}\over
{\partial q}} + q\psi\right )\, dq = \int \left (-{{\partial A}\over
{\partial q}} + qA\right ) \psi \, dq.  \eqno (2.9)$$
The first of these conditions can be satisfied if $A(q, Q) = f(q)
\delta (q - Q)$ [or \hfill \break $f(Q) \delta (q - Q)$],
and the second gives
$A(q,Q) = c e^{-q^2/2} \delta (q-Q)$, $c =$ const.

If one now writes the Hamiltonian for the harmonic oscillator in
terms of $\hat P$ and $\hat Q$ we find (for a convenient factor
ordering)
$$\hat H \Psi = -{{1}\over {2}} \hat P^2 \Psi + \hat Q \hat P \Psi +
{{1}\over {2}} \Psi. \eqno (2.10)$$
The time-independent Schr\"odinger equation becomes
$$-{{1}\over {2}} {{\partial^2 \Psi}\over {\partial Q^2}} +
Q{{\partial \Psi}\over {\partial Q}} + {{1}\over {2}} \Psi = E \Psi,
\eqno (2.11)$$
which, for $E = n + {{1}\over {2}}$ is Hermite's equation with
solutions $\Psi = H_n (Q)$, where the $H_n$ are the Hermite
polynomials.   It is now obvious that $\rho$ is just the usual
convergence factor for the Hermite polynomials
and $A(q,Q)$ maps the harmonic oscillator wave functions to the
Hermite polynomials.  Notice that the (unnormalized) ground state
``Ashtekar'' wave function $\Psi_0 (Q)$ is just one, and that it is
the mapping of $\psi_0 = e^{-q^2/2}$.

The ``Bargmannization'' given here is relatively trivial, and the
mapping
between $\psi$ and $\Psi$ can be achieved by treating the
transformation $Q = q$, $P = q + ip$ as a simple complex canonical
transformation (similar to the real canonical transformations that
Anderson$^{10}$ has been using to good effect lately in quantum
gravity).
The action
$$I = \int [p\dot q - {{1}\over {2}}\{ p^2 + q^2\}]\, dt \eqno
(2.12)$$
becomes
$$I = \int [(P/i)\dot Q + i(q^2/2)^{{\bf \cdot}} - {{1}\over {2}}\{ P^2 +
2PQ\}]\, dt \eqno (2.13)$$
The $(P/i) \dot Q$ term allows one to realize $-i \hat P$ as $-i
\partial /\partial Q$, which gives the previous form of $\hat P$, and
the generating function $G(q)$ of this transformation is $iq^2/2$.
The usual mapping from $\Psi (Q)$ to $\psi (q)$ is $e^{iG}\Psi (Q) =
\psi (q) = e^{-q^2/2} \Psi (Q)$, which was what was found before.
Again notice that for the ground state $\Psi (Q) = 1$, we recover the
usual ground state $\psi (q) = e^{-q^2/2}$.

There is one last element that is needed in order to discuss the
construction of ground state wave functions which is the use of the
``Euclidean'' Hamilton-Jacobi equation ({\it i.e.\/} with the sign of
the potential term reversed).  For the harmonic oscillator discussed
above this equation is
$$ {{1}\over {2}} \left ({{\partial S}\over {\partial q}}\right )^2 -
{{1}\over {2}} q^2 = {{\partial S}\over {\partial t}}.  \eqno
(2.14)$$
In order to find a ground state one begins by taking $\partial
S/\partial t = 0$ and solving (2.14) for S.  Here the solution is
$$ S = \pm q^2 /2 + c^{\prime}, \qquad c^{\prime} = {\rm const.}.
\eqno (2.15)$$
The ground state wave function is then $\psi (q) = e^{-S}$ for the
plus sign in (2.15), or $\psi = ce^{-q^2/2}$, the same function that
was found in the ``Ashtekar'' case.  The main question is how general
this coincidence of the ground states is, especially for gravity.  One
important feature of the Hamilton-Jacobi procedure that will be seen
to be important below is that a partial differential equation such as
(2.14) can have a fairly large number of solutions, even though in this
simple case the only freedom one has is the choice of $c^{\prime}$.

In order to see what happens in quantum gravity I will give a fairly
brief discussion of the application of the two procedures discussed
above to quantum cosmologies.
\vskip 20 pt
\noindent {\bf 3. ``Ground State'' Wave Functions in Quantum
Cosmology}
\vskip 10 pt

A reasonable template for quantum cosmology is the diagonal Bianchi
IX model,
$$ ds^2 = -dt^2 + e^{2\alpha (t)} e^{2\beta (t)}_{ij} \sigma ^i
\sigma ^j, \eqno (3.1)$$
where $\beta_{ij} (t) = {\rm diag} \{\beta_{+} (t) + \sqrt{3}
\beta_{-} (t), \beta_{+} (t) - \sqrt{3} \beta_{-} (t), -2\beta_{+}
(t)\}$ and the $\sigma^i$ are invariant one-forms on the three-sphere
that obey $d\sigma^i = {{1}\over {2}} \varepsilon^{ijk} \sigma^j
\wedge \sigma^k$.  The Ashtekar formalism for this (and other Bianchi
models) is discussed in Refs. [4][5][6].  I will give a quick precis of
the discussion in Ref.[6] (which was based on an article by Friedman
and Jack$^{11}$).  The $\sigma^i$ have the form $\sigma^i \equiv \hat
e^i_a (x^b) dx^a$, where the $\hat e^i_a$ are known non-dynamical
functions of the space coordinates.  The orthonormal one-forms for
the metric (3.1) are $\omega^i = e^{\alpha} e^{\beta}_{ij} \hat e^j_a
dx^a$ $\equiv e^i_a dx^a$.  The spin connections $\Gamma^i_a =
{{1}\over {2}} \varepsilon^{ijk} \Gamma ^j_{ka}$, where $\nabla_a
e^b_j = \Gamma^i_{ja} e^b_i$, are
$$\Gamma^i_a = ({{1}\over {2}}e^{2\beta}_{kk} e^{\beta}_{in} -
e^{3\beta}_{in})\hat e^n_a.  \eqno (3.2)$$
The Ashtekar variables $A^i_a$ are
$$ A^i_a = \Gamma^i_a + i K^i_a, \eqno (3.3)$$
where the $K^i_a$ are $e^{ib}K_{ba}$, $K_{ab}$ the second
fundamental form of $g_{ab} = g_{ij} \hat e^i_a \hat e^j_b$.  It is
easy to show that $K^i_a = P_{in}(t)\hat e^n_a$, so
$$ A^i_a = [G_{in}(t) + iP_{in}(t)] \hat e^n_a \equiv \tau_{in} \hat
e^n_a. \eqno (3.4)$$
The densitized basis elements $\tilde e^a_i$ are $\sqrt{\tilde h}
e^{3\alpha} e^{-\beta}_{ik} \hat e^a_k$, where $\tilde h = {\rm det}
\tilde h_{ab}$, $\tilde h_{ab} = \hat e^i_a \hat e_{ib}$.

Finally, the Hamiltonian constraint, ${\cal H} + i \nabla_a G^a =
-\tilde e^{ia} e^{jb} F^{ij}_{ab}$, where
$$F^{ij}_{ab} = (\varepsilon^{ijk} \tau_{kn} \varepsilon_{n\ell m}
+ \tau_{jm} \tau_{i\ell} - \tau_{im} \tau_{j\ell}) \hat e^m_a \hat
e^{\ell}_b, \eqno (3.5)$$
becomes (after dropping the non-dynamical known functions $\hat
e^i_a$ and $\sqrt{\tilde h}$)
$${\cal H} + i\nabla_a G^a = e^{3\alpha} (2e^{\alpha} e^{\beta}_{ij}
\tau_{ij} - e^{-2\alpha} e^{-\beta}_{kp} e^{-\beta}_{js} \tau_{jp}
\tau_{is} + e^{-2\alpha} e^{-\beta}_{ip} e^{-\beta}_{js} \tau_{ip}
\tau_{js} ). \eqno (3.6)$$
It is now possible to convert ${\cal H} + i \nabla_aG^a$ into an
operator and apply it to a state function $\Psi_A$.  It has become
customary in Ashtekar variable studies to use the ``momentum''
representation where the $A^i_a$ are taken to be coordinates, but I
will use the``coordinate'' representation of Friedman and Jack where
$A^i_a = {{1}\over {2}} \delta /\delta \tilde e^a_i$ and the $\tilde
e^j_i = e^{2\alpha} e^{-\beta}_{ij}$ are the ``coordinates''.  The
$\tilde e^j_i$ are
$$ \tilde e^1_1 \equiv \mu,\qquad \tilde e^2_2 \equiv \nu, \qquad \tilde
e^3_3 \equiv \lambda, \eqno (3.7)$$
where $\mu = e^{2\alpha} e^{-\beta_{+} - \sqrt{3} \beta_{-}}$, $\nu =
e^{2\alpha} e^{-\beta_{+} + \sqrt{3} \beta_{-}}$, $\lambda =
e^{2\alpha} e^{2\beta_{+}}$.  The $\tau_{ij}$ now become
$(1/2)\partial /\partial \tilde e^j_i$, so
$$ \tau_{11} \rightarrow {{1}\over {2}} \partial /\partial \mu,
\qquad \tau_{22} \rightarrow {{1}\over {2}} \partial /\partial \nu ,
\qquad \tau_{33} \rightarrow {{1}\over {2}} \partial / \partial
\lambda. \eqno (3.8)$$
The state function $\Psi_A$ will now be $\Psi_A (\mu, \nu, \lambda)$,
and the Hamiltonian constraint operator applied to $\Psi_A$ gives the
Ashtekar-Wheeler-DeWitt equation:
$$ \nu \lambda {{\partial \Psi_A}\over {\partial \mu}} + \mu \lambda
{{\partial \Psi_A}\over {\partial \nu}} + \mu \lambda {{\partial
\Psi_A}\over {\partial \lambda}} + {{1}\over {2\sqrt{\mu \nu
\lambda}}} \left [ \mu \nu {{\partial^2 \Psi_A}\over {\partial \mu
\partial \nu}} + \nu \lambda {{\partial^2 \Psi_A}\over {\partial \nu
\partial \lambda}} + \mu \lambda {{\partial^2 \Psi_A}\over {\partial
\mu \partial \lambda}} \right ] = 0, \eqno (3.9)$$
where the factors have been ordered so that the derivatives stand to
the right.

This equation is the equivalent of the Schr\"odinger equation (2.11),
where the dreibein variables $\mu$, $\nu$, $\lambda$ play the role
of the coordinate variable $Q$ there.  As for Eq. (2.11), there should be
a ``ground state'' Ashtekar wave function equivalent to $H_0 (Q) =
1$, that is, $\Psi_A =$ const.  There should also exist a mapping,
constructed either by means of a Bargmann-like kernel or by the
canonical transformation procedure.  Notice that the variable $A^i_a
= \Gamma^i_a + iK^i_a$ plays the role of $P = q + ip$ in (2.5), since
the $\Gamma^i_a$ are combinations of the ``coordinate'' variables
$\tilde e^a_i$ (and, of course, their spatial derivatives) and the
extrinsic curvature, $K^i_a$, is basically their conjugate momenta.
One would expect a mapping of the form $\psi_{ADM} = e^{iG} \Psi_A$,
where $\psi_{ADM}$ is a solution to the ordinary Wheeler-DeWitt
equation in ADM variables and $\Psi_A$ is a solution to the
Ashtekar-Wheeler-DeWitt equation (3.9).  Such a mapping was found by
Kodama$^{12}$ and has the form
$$ G = \pm 2i\int \tilde e^a_i \Gamma^i_a \, d^3 x.  \eqno (3.10)$$
Notice that I have used $\tilde e^a_i$ as the coordinate variables
and $A^i_a$ as the $q + ip$ momentum variables.  This means that $G$
is appropriate only for this choice, and that if one were to take
$A^i_a$ as a coordinate variable, as is often done, the $G$ given
here would no longer be appropriate$^{13}$.
In Ref. [6], $G$ was calculated for the diagonal Bianchi Type IX model
and found to be
$$G = i16\pi^2 e^{2\alpha} [e^{-4\beta_{+}} + 2e^{2\beta_{+}}
\cosh (2\sqrt{3} \beta_{-} )]. \eqno (3.11)$$
One would expect the ADM wave function $\psi_{ADM} = e^{iG} \Psi_A$,
where the solution $\Psi_A = 1$ is taken, to be a ``ground state''
wave function for the diagonal Bianchi IX model.  In fact, in Ref. [6]
we showed that there is a close connection between this ``Ashtekar''
solution and one constructed from the ``Euclidean''
Einstein-Hamilton-Jacobi equation for the
diagonal Bianchi IX models,
$$ \left ({{\partial S}\over {\partial \alpha}} \right )^2 - \left
({{\partial S}\over {\partial \beta_{+}}}\right )^2 - \left
({{\partial S}\over {\partial \beta_{-}}} \right )^2 + e^{4\alpha}
V(\beta_{\pm}) = 0, \eqno (3.12)$$
where the ``three-curvature'' term $e^{4\alpha} V(\beta_{\pm})$ has
the opposite sign from the term in the usual
Einstein-Hamilton-Jacobi equation and the potential $V(\beta_{\pm})$
is the usual $V(\beta_{\pm}) = {{1}\over {3}} e^{-8\beta_{+}} -
{{4}\over {3}}e^{-2\beta_{+}} \cosh (2\sqrt{3} \beta_{-}) + {{2}\over
{3}}e^{4\beta_{+}} [\cosh (4\sqrt{3} \beta_{-}) - 1]$.  One solution
to (3.12) is $S = {{1}\over {6}} e^{2\alpha} [e^{-4\beta_{+}} +
2e^{2\beta_{+}} \cosh (2\sqrt{3} \beta_{-} )]$, so $e^{-S}$ is $\exp
[(3/8\pi^2) iG]$.  Because (3.12) is nonlinear, the constant factor in
the exponent by which these two expressions differ cannot be removed
by a simple rescaling of $S$.  However, Eq. (3.1) implies a choice of
normalization constant in the definition of the three metric, that
is, Eq. (3.1) should read $ds^2 = -dt^2 + R_0^2 \exp(2\alpha)
\exp(2\beta_{ij}) \sigma^i \sigma^j$, and $R_0$ was taken to be
one.  By choosing $R_0$ properly, the two expressions for the
``ground state'' can be made to coincide.  Note, however, that these
two wave functions {\it do\/} differ, even if the difference is trivial and
removable.

It seems that we have discovered a ``magic bullet'' that can generate
``ground state'' wave functions, since the expression for $G$ in Eq.
(3.10) is constructive and can be calculated for any metric, so $e^{iG}$
can be generated easily, while the equivalent of Eq. (3.12) is a
nonlinear functional differential equation that must be solved.
Unfortunately, as I will show below, $e^{iG}$ does not always
correspond to interesting ``ground state'' solutions.  In the next
section of the article I will look at the electromagnetic analogue of the
gravitational problem.

The $e^{iG}$ ``ground state'' for the Bianchi Type I models is easy
to calculate since the spatial metric is
$$d\sigma^2 = e^{2\alpha} e^{2\beta}_{ab} dx^a dx^b, \eqno (3.15)$$
the dreibein vectors $\hat e^i_a$ are just $\hat e^i_a =
\delta^i_a$, and the orthonormal vectors are $e^{\alpha}
e^{\beta}_{ij} \delta^i_a$.  Since both the metric $h_{ab} = e^i_a
e_{ib} = e^{2\alpha} e^{2\beta}_{ab}$ and the $e^i_a$ themselves are
constant on $t =$ const. hypersurfaces, $\nabla_a e^b_j = 0$ and
$\Gamma^i_a = 0$.  This means that $\tilde e^a_i \Gamma^i_a = 0$ and
$G = 0$.  The $e^{iG}$ solution is just equal to one.  The
Einstein-Hamilton-Jacobi equation in this case is
$$\left ({{\partial S}\over {\partial \alpha}} \right )^2 - \left
({{\partial S}\over {\partial \beta_{+}}} \right )^2 - \left
({{\partial S}\over {\partial \beta_{-}}} \right )^2 = 0, \eqno
(3.16)$$
and $S = 0$ is, of course, a solution, so $e^{-S} = 1$ gives the same
state function as above, but whether or not this solution is a
``ground state'' is a moot point.  There is, however, another system
for which the $e^{iG}$ state probably has even less claim to be a
``ground state'' than in the Bianchi I case.  This system is the
polarized Gowdy metric$^{14}$,
$$ds^2 = e^{t-\lambda/2} (-e^{4t}dt^2 + d\theta^2) + e^{2t}
(e^{\beta}d\sigma^2 + e^{-\beta}d\delta^2), \eqno (3.17)$$
where $\beta = \beta(\theta, t)$, $\lambda = \lambda (\theta, t))$,
$0 \leq \theta, \sigma, \delta < 2\pi$.  The Hamiltonian formulation
and the quantization of this model were first discussed by
Misner$^{15}$
and Berger$^{16}$.  A full discussion of the ``ground state'' problem
will be given elsewhere$^{17}$, but I will give a quick sketch of the
results to show where the problem is with taking $e^{iG}$ to be the
``ground state''.  The Hamiltonian constraint for this metric is
$${\cal H} = H + {{1}\over {2\pi}}\int_0^{2\pi} d\theta [{{1}\over
{2}} p^2_{\beta} + {{1}\over {2}} e^{4t} (\beta^{\prime})^2] = 0,
\eqno (3.18)$$
where $\prime$ is $\partial /\partial \theta$, $p_{\beta}$ is the
momentum conjugate to $\beta$, and $H = (1/2\pi) \int_0^{2\pi} \, d\theta
p_t$, where $p_t$ is the momentum conjugate to $t$.  It is obvious
that a gauge has been chosen where the metric component $t$ is an
internal time, so $H$ is the ADM Hamiltonian of the problem.  Since
$H$ plays the role of the Hamiltonian, the ``Euclidean''
Einstein-Hamilton-Jacobi equation is
$$ {{\partial S}\over {\partial t}} + {{1}\over {2\pi}}\int_0^{2\pi}
d\theta \left[ {{1}\over {2}}\left ({{\delta S}\over {\delta
\beta}}\right )^2 - {{1}\over {2}}e^{4t} (\beta^{\prime})^2 \right ]
= 0. \eqno (3.19)$$
If one expands $\beta$ and $p_{\beta}$ in real Fourier series,
$$\beta = q_0 + \sqrt{2} \sum_{n=1}^{\infty} [q_n(t) \cos n\theta +
q_{-n}(t) \sin n\theta ], $$
$$ p_{\beta} = p_0 + \sqrt{2} \sum_{n=1}^{\infty} [p_n(t) \cos
n\theta + p_{-n}(t) \sin n\theta ], \eqno (3.20)$$
Misner and Berger have shown that the Wheeler-DeWitt equation is
$$i{{\partial \psi}\over {\partial t}} = \sum_{n = -\infty}^{\infty}
{{1}\over {2}} \left ( -{{\partial^2 \psi}\over {\partial q_n^2}} + n^2
e^{4t} q_n^2 \psi \right ). \eqno (3.21)$$
The ``ground state'' solution of Berger$^{16}$ is
$$\psi_{GS} = \exp(-\sum_{n} A_n(t) q_n^2), \eqno (3.22)$$
where the $A_n(t)$ are given by Berger.  The relation of this ground
state to the $e^{-S}$ state will be considered in Ref. [17].

The Ashtekar ``ground state'' $e^{iG}$ is readily calculated.  If one
takes the orthonormal basis $\sigma^1 = e^{t/3 - \lambda/4} d\theta$,
$\sigma^2 = e^{t + \beta/2} d\sigma$, $\sigma^3 = e^{t - \beta/2}
d\delta$, the basis vectors $e^i_a$ are diagonal ({\it i.e.\/} $e^i_a
= f_a \delta^i_a$), while the only non-zero $\Gamma^i_a$ are the
non-diagonal terms $\Gamma^2_3$ and $\Gamma^3_2$, so $e^a_1
\Gamma^i_a = 0$, and, as in the Bianchi I case the Ashtekar ``ground
state'' is $\Psi = 1$.  The is obviously unrelated to the state given
in Eq. (3.22).  Of course, the fact that $G$ is not gauge invariant
means that one can rotate the basis used here and make $G$ into
anything desired, but this seems a bit artificial if the Ashtekar
``ground state'' is to be something that comes from a constructive
procedure that is easy to implement.

The fact that the Ashtekar variable procedure leads to ``ground
states'' that are not what one would expect, and especially to states
that seem to differ from those obtained from the
Einstein-Hamilton-Jacobi procedure, means that there must be a
fundamental difference between the two procedures.  In the next
section I will consider the simpler system of a free electromagnetic
field, where the differences between the analogues of the two
procedures is more easily understood, and the coincidence between
them in certain cases is more understandable.
\vskip 20 pt
\noindent {\bf 4. Electromagnetic ``Ground State'' Wave Functions}
\vskip 10 pt

If one considers the usual action for the free electromagnetic field
in Hamiltonian form,
$$I = {{1}\over {16\pi}} \int [\pi^i \dot A_i - \{{{1}\over {8}}\pi^i
\pi^i +  2 (\nabla \times {\bf A})^2 \} - A_0 \pi^i_{,i} ]\, d^4 x,
\eqno (4.1)$$
$i = 1,2,3$ and expands $A_0$, the $A_i$ and their conjugate momenta
$\pi^i$ in real Fourier series (using box normalization in a box of
side $L$) as
$$A_i = {{1}\over {L^{3/2}}} \sum_{\n} \left [A_i^{(\n)}(t) \cos
\left ({{2\pi}\over {L}} \n\cdot {\bf x}\right)
+ \tilde A_i^{(\n)} (t) \sin \left ( {{2\pi}\over {L}} \n \cdot {\bf
x} \right ) \right ],$$
$$\pi^i = {{1}\over {L^{3/2}}}\sum_{\n} \left [ P^{i(\n)}(t) \cos \left
({{2\pi}\over {L}} \n \cdot {\bf x} \right ) + \tilde P^{i(\n)}(t)
\sin \left ({{2\pi}\over {L}} \n \cdot {\bf x} \right )\right ],$$
$$A_0 = {{1}\over {L^{3/2}}}\sum_{\n} \left [ A_0^{(\n)}(t) \cos \left
({{2\pi}\over {L}} \n \cdot {\bf x} \right ) + \tilde A_0^{(\n)}(t)
\sin \left ({{2\pi}\over {L}} \n \cdot {\bf x} \right )\right ], \eqno
(4.2)$$
the action reduces to
$$ I = {{1}\over {32\pi}} \int \sum_{\n} [ P^{i(\n)} \dot A_i^{(\n)}
+ \tilde P^{i(\n)} \dot {\tilde A}_i^{(\n)} -$$
$$ - \{ {{1}\over {8}} P^{i(\n)} P^{i(\n)} + {{1}\over {8}} \tilde
P^{i(\n)} \tilde P^{i(\n)} + {{8\pi^2}\over {L^2}} (\n \times {\bf
A}^{(\n)})^2 +$$
$$ + {{8\pi^2}\over {L^2}} (\n \times {\tilde {\bf A}}^{(\n)})^2 \} -
A_0^{(\n)} \n \cdot {\bf P}^{(\n)} - \tilde A_0^{(\n)} \n \cdot
\tilde {\bf P}^{(\n)} ] dt. \eqno (4.3)$$

If the sign of the potential term in the above action is changed, the
``Euclidean'' Hamilton-Jacobi equation that corresponds to the action
is
$$ {{\partial S}\over {\partial t}} = \sum_{\n} \bigg [ {{1}\over
{8}} {{\partial S}\over {\partial A_i^{(\n)}}}{{\partial S}
\over {\partial A_i^{(\n)}}}  +  {{1}\over {8}} {{\partial S}\over
{\partial \tilde A_i^{(\n)}}} {{\partial S}\over {\partial \tilde
A_i^{(\n)}}} -$$
$$- {{8\pi^2}\over {L^2}} (\n \times {\bf A}^{(\n)})^2 -
{{8\pi^2}\over {L^2}} (\n \times {\bf A}^{(\n)} )^2 \bigg ]. \eqno
(4.4)$$
For $\partial S /\partial t = 0$ we can find two distinct solutions
to this equation.  For simplicity I will take the Coulomb gauge, $\n
\cdot {\bf A}^{(\n)} = \n \cdot \tilde {\bf A}^{(\n)} = 0$.  The first of
these solutions is
$$S = - \sum_{\n} {{8\pi}\over {L}} |\n| \{ |{\bf A}^{(\n)} |^2 +
|\tilde {\bf A}^{(\n)} |^2 \}.  \eqno (4.5)$$
The function $e^{S}$ is the box equivalent of the Wheeler ground
state functional for the free electromagnetic field$^{18}$,
$$ \psi = {\cal N} \exp \left \{ -\int \int {{{\bf B} ({\bf x}_1 )
\cdot {\bf B} ({\bf x}_2)}\over {16\pi^3 \hbar c r^2_{12}}} d^3 x_1
d^3 x_2 \right \}. \eqno (4.6)$$
However, there is another solution for $\partial S /\partial t = 0$
that can be obtained by rewriting the right-hand side of (4.4), that is,
$$ \sum_{\n}  \bigg \{ \left [ {{\partial S}\over {\partial
A^{(\n)}_i}} + {{8\pi}\over {L}} (\n \times \tilde {\bf A}^{(\n)}
)\right ] \left [{{\partial S}\over {\partial A^{(\n)}_i}} -
{{8\pi}\over {L}} (\n \times \tilde {\bf A}^{(\n)} ) \right ] +$$
$$ + \left [ {{\partial S}\over {\partial \tilde A^{(\n)}_i}} +
{{8\pi}\over {L}} (\n \times {\bf A}^{(\n)} ) \right ] \left [
{{\partial S}\over {\partial \tilde A^{(\n)}_i}} - {{8\pi}\over {L}} (\n
\times {\bf A}^{(\n)} ) \right ]\bigg \} = 0. \eqno (4.7)$$
This form of the equation shows that
$$ S = \pm {{4\pi}\over {L}} \sum_{\n} \left [ (\n \times \tilde {\bf
A}^{(\n)} ) \cdot {\bf A}^{(\n)}  - \tilde {\bf A}^{(\n)} \cdot (\n
\times {\bf A}^{(\n)} ) \right ] \eqno (4.8)$$
is also a solution.  This solution is
$$ S = \mp 4 \int {\bf A} \cdot {\bf B} \, d^3 x = \mp 4 \int {\bf A}
\cdot (\nabla \times {\bf A} ) \, d^3 x, \eqno (4.9)$$
which is the Chern-Simons term for a $U(1)$ gauge theory.

If one takes $e^{-S}$ for (4.6) and (4.9) one can generate two ``ground
state'' quantum solutions.  Which of these two corresponds to the
$e^{iG}$ Ashtekar solution of the previous section?  It is possible
to set up the analogue of the Ashtekar variables for the free
electromagnetic field by taking the new variables $\pi^{\prime i} =
\pi^i + 4B^i$ and $A^{\prime}_i = A_i$.  The mode parameters of
$\pi^{\prime i}$ are
$$P^{\prime i (\n)} = iP^{i (\n)} + {{8\pi}\over {L}} (\n \times
\tilde {\bf A}^{(\n)})^i,$$
$$\tilde P^{\prime i (\n)} = i \tilde P^{i (\n)} - {{8\pi}\over {L}}
(\n \times {\bf A}^{(\n)})^i. \eqno (4.10)$$
The Hamiltonian in terms of these variables (with all momenta
standing to the right) is
$$H = \sum_{\n} \bigg [-{{1}\over {8}} P^{\prime i (\n)} P^{\prime i
(\n)} + {{2\pi}\over {L}} (\n \times \tilde {\bf A}^{(\n)} )^i
P^{\prime i (\n)} -$$
$$-{{1}\over {8}} \tilde P^{\prime i (\n)} \tilde
P^{\prime i (\n)} - {{2\pi}\over {L}} (\n \times {\bf A}^{(\n)} )^i
\tilde P^{\prime i (\n)} \bigg ]. \eqno (4.11)$$
If the $P^{\prime i (\n)}$ and $\tilde P^{\prime i (\n)}$ are
realized as $\partial /\partial A_i^{(\n)}$ and $\partial /\partial
\tilde A_i^{(\n)}$, $\hat H \Psi (A_i^{(\n)}, \tilde A_i^{(\n)}, t) =
i\partial \Psi/ \partial t$ has as a time-independent solution $\Psi
=$ const.  The generating function of the transformation between
$P^{i(\n)}$, $A^{(\n)}_i$, $\tilde P^{i(\n)}$, $\tilde A^{(\n)}_i$
and $P^{\prime i(\n)}$, $A^{(\n)}_i$, $\tilde P^{\prime i(\n)}$,
$\tilde A^{(\n)}_i$ is calculated in exactly the same way as for the
harmonic oscillator example in Sec. 2, and we have
$$ G(A^{(\n)}_i, \tilde A^{(\n)}_i ) = {{4\pi i}\over {L}} \sum_{\n}
\left [(\n \times \tilde {\bf A}^{(\n)} )\cdot {\bf A}^{(\n)} +
\tilde {\bf A}^{(\n)} \cdot (\n \times {\bf A}^{(\n)} ) \right ]$$
$$= 4i\int {\bf A} \cdot (\nabla \times {\bf A} )\, d^3 x. \eqno
(4.12)$$
So, $e^{iG}$ is $e^{-S}$ for the Chern-Simons $S$ given by (4.9).

The two solutions (4.6) and (4.9) are both valid solutions for $S$, but
only (4.6) is the true ground state, while $e^{-S}$ for the proper sign
of (4.9) and the ``Ashtekar'' $e^{iG}$ cannot reasonably be called a
``ground state'' of the free electromagnetic field.  The
electromagnetic ``Ashtekar'' analysis leads one to suspect that the
Ashtekar ``ground state'' for the diagonal Bianchi IX model given in
the last section is somehow the equivalent of (4.9), but, if so, the
striking coincidence between it and the solution from the
Einstein-Hamilton-Jacobi equation should be explained.  Of course, it
might be that the gravitational Ashtekar state and the ``ground
state'' solution from the Einstein-Hamilton-Jacobi equation are
simply the same, but a study of the ``cosmological'' equivalents of
the free electromagnetic field shows that the coincidence is most
likely an artifact of homogeneity of the cosmological gravitational
field.

By ``cosmological'' electromagnetic field I mean a vector potential
defined as a spatially homogeneous vector field over a Bianchi-type
three-space.  That is, ${\bf A} = A_i(t) \sigma^i$, where the
$\sigma^i = \hat e^i_a dx^a$ are homogeneous one-forms on one of the
nine Bianchi three-spaces.  Here $d\sigma^i = (1/2)C^i_{jk} \sigma^j
\wedge \sigma^k$ [or $(\hat e^i_{a,b} - \hat e^i_{b,a})\hat e^b_j \hat
e^a_k = C^i_{jk} ]$, $C^i_{jk} =$ const.  The three-space metric is
no longer Euclidean, so indices must be raised and lowered correctly
and, in principle, covariant derivatives must be used, but the curl
of $A_a$ is unchanged, and $g^{ac} g^{bd} (A_{c,d} - A_{d,c})
(A_{a,b} - A_{b,a})$ for example is $C^i_{\ell m} C^k_{\ell m} A_i
A_k$.  The action for the electromagnetic field is ($\pi^i (t) =
\pi^a \hat e^i_a$, $\pi^i_{,i} =  \hat e^a_i \pi^i_{,a} = 0$)
$$I = {{1}\over {16\pi}} \int \{ [ \pi^i \dot A_i - ({{1}\over {8}}
\pi^i \pi^i + C^i_{\ell m} C^k_{\ell m} A_iA_k)]\sigma^1 \wedge
\sigma^2 \wedge \sigma^3 \} dt. \eqno (4.14)$$
For all class A Bianchi models $C^i_{jk} = \varepsilon_{\ell
jk}m^{\ell i}$, $m^{ij} =$ const., so $C^i_{\ell m}C^k_{\ell m} = 2
m^{is}m^{sk}$, and
$$ I = {{V}\over {16\pi}} \int \{ \pi^i \dot A_i - ({{1}\over {8}}
\pi^i \pi^i + 2 m^{is} m^{sk} A_iA_k)\}dt, \eqno (4.15)$$
where $V \equiv \int \sigma^1 \wedge \sigma^2 \wedge \sigma^3$ (and
the space is artificially closed when necessary).

{}From the Hamiltonian $(1/8)\pi^i \pi^i + 2m^{is}m^{sk} A_iA_k$, the
Euclidean Hamilton-Jacobi equation is
$${{\partial S}\over {\partial t}} = {{1}\over {8}} {{\partial
S}\over {\partial A_i}} {{\partial S}\over {\partial A_i}} -
2m^{is}m^{sk}A_iA_k. \eqno (4.16)$$
For $\partial S/\partial t = 0$ the ``ground state'' solution is $S =
\pm 2m^{ij} A_iA_j$.  Also for $\partial S /\partial t = 0$ it is
possible to factor Eq. (4.16) as was done in Eq. (4.7) as
$$\left ({{\partial S}\over {\partial A_i}} + 4m^{is}A_s \right )
\left ({{\partial S}\over {\partial A_i}} - 4m^{ik}A_k \right ) = 0.
\eqno (4.17)$$
This equation has two solutions which are exactly the same $S = \pm
2m^{ij}A_iA_j$ as the ``ground state'' solution given above.  One can
also make the Ashtekar transformation $\pi^{\prime i} = i\pi^i + 4
m^{is}A_s$, $A^{\prime}_i = A_i$, and the Hamiltonian reduces to
$$-{{1}\over {8}} \pi^{\prime i} \pi^{\prime i} + m^{is}A^{\prime}_s
\pi^{\prime i}. \eqno (4.18)$$
If in the quantum theory one realizes $\hat \pi^{\prime i}$ as
$\partial /\partial A_i$, $\hat H \Psi = 0$ has as a solution $\Psi
=$ const.  It is easy to show that the generating function of the
transformation is $G = i(V/16\pi) (2m^{is}A_iA_s)$.

In this cosmological case the three methods of generating a ``ground
state'' wave function, $e^{-S}$ for the ``Wheeler'' $S$, the
Chern-Simons $S$, and the ``Ashtekar'' $e^{iG}$ give essentially the
same function.  Notice that $G$ has the overall factor $V/16\pi$,
which is a distinguishing feature noted in Ref. [6].  Of course, the
three space can be rescaled in such a way that $V = 16\pi$, as was
done there, so the coincidence between the three functions becomes
exact.  Notice also that for the Bianchi I space all three wave
functions are constant.

It is easy to see that the reason the three different methods lead to
the same wave function is that the curl of ${\bf A}$ is a constant
matrix times ${\bf A}$ in this ``cosmological'' case.  In general,
$\nabla \times {\bf A}$ can be quite different from ${\bf A}$ itself.
For a constant ${\bf B}$ field in flat space, ${\bf A} = {\bf B}
\times {\bf r}$, which is perpendicular to ${\bf B}$, so ${\bf A}
\cdot {\bf B} = 0$.  However, for a homogeneous field over a Class A
Bianchi space (with the exception of the Bianchi I space), the linear
relation between $\nabla \times {\bf A}$ and ${\bf A}$ means that the
Chern-Simons (and ``Ashtekar'') wave functions are equal to the
``Wheeler'' ground state.  Unfortunately, this is fortuitous and the
Chern-Simons state cannot, in general, be the ground state of the
free quantum electromagnetic field.
\vskip 20 pt
\noindent {\bf 5. Conclusions}
\vskip 10 pt

In Ref. [6] the Ashtekar variable approach was used to generate a wave
function of the Bianchi-type IX universe that was the same as one
constructed from a solution to the ``Euclidean''
Einstein-Hamilton-Jacobi equation. In the present article it was
shown that the coincidence of these two solutions in that case was
something of an accident, since the Ashtekar solution is closer
in spirit to one constructed from the Chern-Simons term in the
Maxwell field.  In general on cannot expect these two solutions to
coincide, and the polarized Gowdy model is an example where they
do not.

Notice, however, that for the free electromagnetic field the
Chern-Simons term generates an acceptable quantum solution for the
field, whether or not it can be called a true ``ground state''.  In
the gravitational case it should be possible to generate these
solutions in a number of cases which will give a set of perhaps
useful solutions to the Wheeler-DeWitt equation.  From the analysis
of the ``cosmological'' electromagnetic fields, one might expect that
the Ashtekar procedure will lead to correct ``ground state''
solutions. at least in the case of Bianchi quantum cosmologies, and
perhaps in other cases as well.
\vskip 20 pt
\noindent {\bf Acknowledgements}
\vskip 10 pt

This article is based in part on discussions with V. Moncrief in
preparation for a more extended article.
\vskip 20 pt
\noindent {\bf References}
\vskip 10 pt

\item {1.} J. Pleba\'nski, {\sl J. Math. Phys.} {\bf 18} (1977) 2511.
\vskip 10 pt
\vfill\eject
\item {2.} See, for example, A. Barajas, G. D. Birkhoff, C. Graef, and
N. Vallarta, {\sl Phys. Rev.} {\bf 66} (1944) 54; M. Moshinsky, {\sl
Phys. Rev.} {\bf 80} (1950) 514.
\vskip 10 pt

\item {3.} See, for example, A. Ashtekar, {\sl New Perspectives in
Quantum Gravity} (Bibliopolis, Naples, 1988).
\vskip 10 pt

\item {4.} A. Ashtekar and J. Pullin, {\sl Ann. Israel Phys. Soc.}
{\bf 9} (1990) 66.
\vskip 10 pt

\item {5.} O. Obregon, J. Pullin, and M. Ryan, To appear, {\sl Phys.
Rev. D}.
\vskip 10 pt

\item {6.} V. Moncrief and M. Ryan, {\sl Phys. Rev.} {\bf D44} (1991)
2375.
\vskip 10 pt

\item {7.} V. Bargmann, {\sl Comm. Pure App. Math.} {\bf 14} (1961)
187.
\vskip 10 pt

\item {8.} C. Misner, K. Thorne, and J. Wheeler, {\sl Gravitation}
(Freeman, San Francisco, 1973).
\vskip 10 pt

\item {9.} V. Fock, {\sl Z. Physik} {\bf 49} (1928) 339.
\vskip 10 pt

\item {10.} A. Anderson, Private communication.
\vskip 10 pt

\item {11.} J. Friedman and I. Jack, {\sl Phys. Rev.} {\bf D37}
(1988) 3495.
\vskip 10 pt

\item {12.} H. Kodama, {\sl Prog. Theor. Phys.} {\bf 80} (1988) 1024;
{\sl Phys. Rev.} {\bf D36} (1987) 1587.
\vskip 10 pt

\item {13.} R. Capovilla, Private communication.
\vskip 10 pt

\item {14.} R. Gowdy, {\sl Phys. Rev. Lett.} {\bf 27} (1971) 826;
{\sl Ann. Phys. (N. Y.)} {\bf 83} (1974) 203.
\vskip 10 pt

\item {15.} C. Misner, {\sl Phys. Rev.} {\bf D8} (1973) 3271.
\vskip 10 pt

\item {16.} B. Berger, {\sl Ann. Phys. (N. Y.)} {\bf 83} (1974) 458;
{\sl Phys. Rev.} {\bf D11} (1975) 2770.
\vskip 10 pt

\item {17.} V. Moncrief and M. Ryan, In preparation.
\vskip 10 pt

\item {18.} J. Wheeler, {\sl Geometrodynamics} (Academic Press, New
York, 1962).

\end